\documentclass[aps,pre,reprint,nofootinbib,groupedaddress]{revtex4-2}
\usepackage{amsmath,amssymb,mathtools,bm,physics}
\usepackage{appendix,titlesec,graphicx,latexsym,setspace,lineno}
\usepackage[pdftex, pdfborder={0 0 0}, colorlinks=true, linkcolor=red,citecolor=blue, urlcolor=blue]{hyperref}
\usepackage[usenames,dvipsnames,svgnames,table]{xcolor}
\usepackage{orcidlink}
\usepackage[capitalise]{cleveref}
\crefname{section}{Sec.}{Secs.}
\crefname{figure}{Fig.}{Figs.}
\crefname{equation}{Eq.}{Eqs.}

\allowdisplaybreaks[3]

\newcommand{\nn}{\nonumber}

\begin{document}
\title{Activity-Enhanced Ordering in Fluctuation-Induced First-Order Transitions}
\author{Suvendra K. Sahoo
\,\orcidlink{0009-0004-6999-7897}
}
\email{suvendrak@iisc.ac.in}
\affiliation{Centre for Condensed Matter Theory, Department of Physics, Indian Institute of Science, Bengaluru 560 012, India}

\begin{abstract}
    Fluctuations can drive otherwise continuous phase transitions to first order through the Brazovskii mechanism.
    We study how these fluctuation-induced transitions are modified in active systems by introducing nonequilibrium spatiotemporally correlated noise.
    We show that, while the transition remains fluctuation-induced first order, activity systematically suppresses these fluctuation effects, shifting the transition to higher temperatures and rendering it increasingly weakly first order.
    As a result, ordering is enhanced without inducing a spinodal instability of the isotropic phase, as confirmed by direct numerical simulations.
    In the strong-activity limit, fluctuation effects disappear and mean-field behavior is recovered.
    Our results identify activity as a generic control parameter for tuning the strength of fluctuation-induced first-order transitions.
\end{abstract}
\maketitle

\section{Introduction}
\label{sec:intro}
\begin{figure}[t]
    \includegraphics[width=\columnwidth]{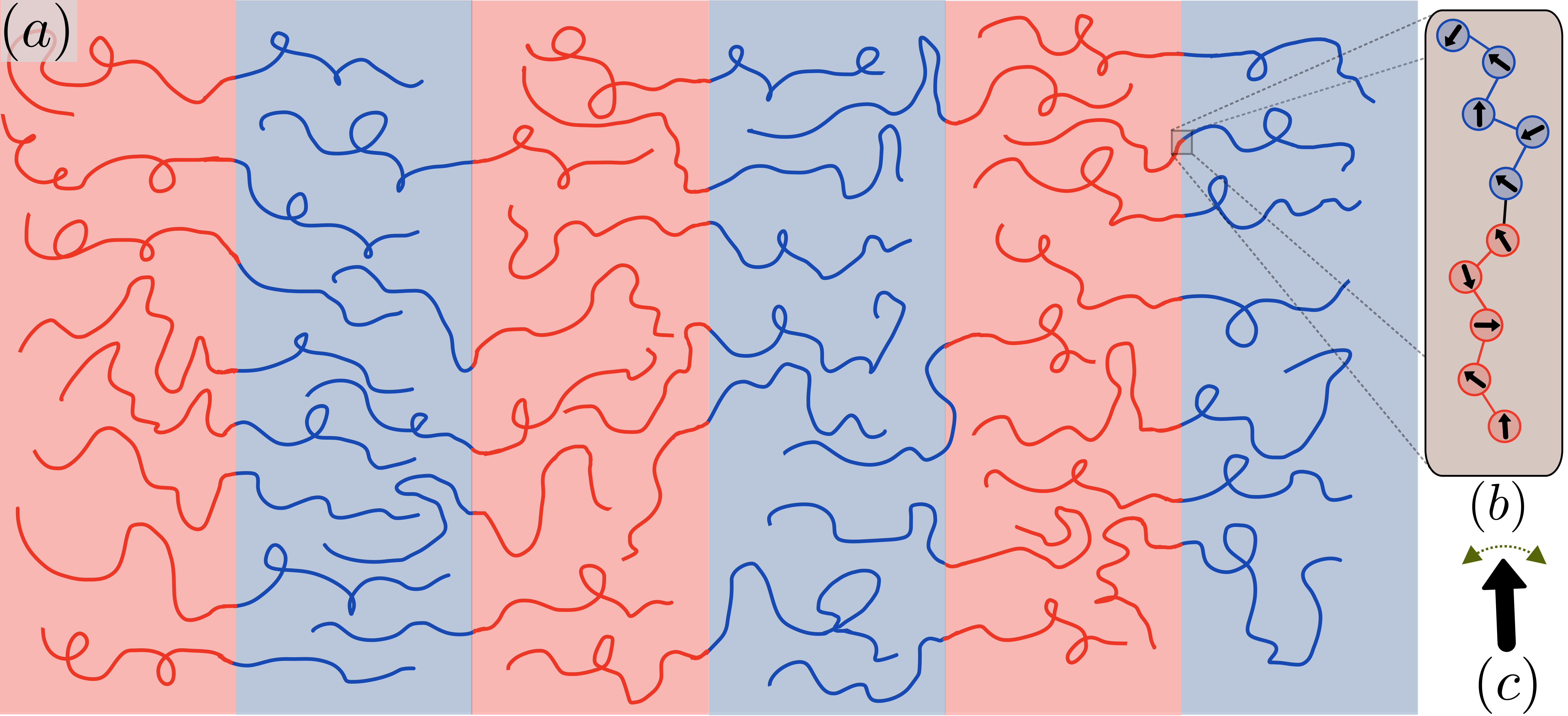}
    \caption{(a) Schematic illustration of active diblock copolymer with equal length subchains exhibiting microphase separation, (b) Red and blue segments represent distinct monomer species; interactions between unlike monomers are energetically unfavorable compared to like-like interactions, (c) Each monomer possesses a self-propulsion velocity, with its direction (indicated by arrows) evolving via diffusive dynamics.}
    \label{diblock}
\end{figure}

Fluctuations can play a fundamental role in determining the character of phase transitions. In systems that order at a finite wavelength, they convert a transition predicted by mean-field theory to be continuous into a weakly first-order one via the Brazovskii mechanism \cite{brazovskii1995PhaseTransition}. This scenario underlies a wide range of pattern-forming systems, including block copolymers \cite{fredrickson1989KineticsMetastable}, liquid crystals near the nematic--smectic C transition \cite{grinstein1986DefectmediatedMelting}, and fluids near the Rayleigh--B\'enard instability \cite{swift1977HydrodynamicFluctuations}, where ordering leads to spatially modulated structures with different symmetries \cite{lavrentovich2016FirstorderPatterning}.
A natural question is how this fluctuation-driven mechanism is modified in active systems. Activity is ubiquitous in biological and synthetic matter and generically drives systems out of equilibrium through the continuous input and dissipation of energy at the constituent level \cite{ramaswamy2017ActiveMatter,marchetti2013HydrodynamicsSoft}.
Here, we address the question by introducing activity in its minimal form as spatiotemporally correlated (colored) noise with finite persistence time $\tau$ \cite{fodor2016HowFar,maitra2020EnhancedOrientational} and correlation length $\xi$ \cite{maggi2022CriticalActive} into the Brazovskii framework.
While the effects of colored noise have been explored in related settings \cite{garcia-ojalvo1993EffectsExternal,garcia-ojalvo1994ColoredNoise,maitra2020EnhancedOrientational}, its role on fluctuation-induced first-order transitions has not been examined.
Interestingly, colored noise is only one route to breaking detailed balance and pre-dates the active-matter literature, e.g., in dye lasers \cite{hanggi1995ColoredNoise,jung1987DynamicalSystems,fox1986UniformConvergence}. 
Coarse-graining self-propelled particles commonly yields deterministic terms that cannot be derived from a free energy, as in Active Model B/B+ \cite{cates2015MotilityInduced,wittkowski2014ScalarField,tjhung2018ClusterPhases} and non-reciprocal Cahn--Hilliard dynamics \cite{saha2020ScalarActive}; whether these deterministic terms or the colored noise arise in the effective field theory can be determined only by coarse-graining from the microscopics, which is beyond our scope here.
Nevertheless, colored-noise field theories have been used to describe active systems, where the noise correlation time and length are set by the microscopic persistence time of the active agents and their velocity-correlation length, respectively
\cite{paoluzzi2016CriticalPhenomena,maitra2020EnhancedOrientational,maggi2022CriticalActive}; thus, we consider the following noise sector as a minimal description of activity.
We discuss possible experimental realizations, e.g., \cref{diblock}, towards the end of the paper.

We examine the approach to stationarity as well as the resulting steady state. Here are our main results. Recall first \cite{gross2000KineticsOrdering} that at early times fluctuations do not modify the mean-field dynamics: quenches below the mean-field transition still exhibit unstable growth toward the ordered (patterned) state. At longer times, however, fluctuations remain essential and qualitatively modify the transition. The steady state is instead controlled by a fluctuation-induced first-order transition, characterized by a metastable disordered phase and ordering via nucleation.
Increasing activity (through increasing $\tau$ or $\xi$) systematically suppresses the late-time fluctuation effects underlying the Brazovskii mechanism, thereby rendering the transition progressively more weakly first-order. As a consequence, the transition shifts to higher temperatures and the ordering is enhanced. In the infinite-activity limit
($\tau,\xi\to\infty$), fluctuation corrections vanish, and the system crosses over to mean-field behavior.

\section{Model}
\label{sec:model}
We now show how these results were obtained. We consider a standard relaxational dynamics for a scalar order parameter $\phi(\mathbf{x},t)$ governed by
\begin{equation}
    \label{phieq}
    \partial_t \phi(\mathbf{x},t) = -D(i\nabla)^a\frac{\delta F[\phi]}{\delta \phi} + \zeta(\mathbf{x},t),
\end{equation}
where the exponent $a=0,\,2$ corresponds to non-conserved (\textit{Model A}) and conserved (\textit{Model B}) dynamics, respectively \cite{hohenberg1977TheoryDynamic}. The free-energy functional is taken to be
\begin{equation}
\label{Fphi}
F[\phi] = \int_{\mathbf{x}}\frac{r_0}{2}\phi^2 + \frac{u}{4!}\phi^4 + \frac{\kappa}{2}\left[(\nabla^2+q_0^2)\phi\right]^2,
\end{equation}
which favors ordering at a finite wave vector $q_0$. This is the Swift--Hohenberg free energy, widely used as a minimal realization of the Brazovskii model, since both exhibit an isotropic fluctuation spectrum of soft modes peaked at $q_0$ \cite{hohenberg1995MetastabilityFluctuationdriven,swift1977HydrodynamicFluctuations}. At the mean-field level, $r_0=0$ marks the critical point, with $r_0<0$ corresponding to a spatially modulated ordered phase.
Note that when the $\phi\to-\phi$ symmetry is absent, as in asymmetric diblock copolymers \cite{leibler1980TheoryMicrophase} or weak crystallization \cite{kats1993WeakCrystallization}, a cubic term is allowed in \cref{Fphi}. As a result, the transition becomes first order already at the mean-field level, while fluctuation corrections stabilize a hexagonal phase at sufficiently large asymmetry \cite{brazovskii1995PhaseTransition}. We restrict to the symmetric case \cite{fredrickson1989KineticsMetastable}, where the cubic term is forbidden and fluctuations are the sole origin of the first-order character.

The stochastic forcing $\zeta(\mathbf{x},t)$ represents fast degrees of freedom and is taken to be Gaussian with zero mean and correlator
\begin{equation}
\left<\zeta(\mathbf{x},t)\zeta(\mathbf{x}',t')\right> = 2D(i\nabla)^a K_{\xi,\tau}(|\mathbf{x}-\mathbf{x}'|,|t-t'|),
\end{equation}
characterized by a correlation length $\xi$ and persistence time $\tau$.
A convenient realization of the kernel $K_{\xi,\tau}$ is obtained by introducing an auxiliary field $\zeta'(\mathbf{x},t)$ obeying an Ornstein--Uhlenbeck (OU) process,
\begin{equation}
    \label{OUeq}
    \tau \partial_t \zeta'(\mathbf{x},t) = -\zeta' + \xi^2 \nabla^2 \zeta' + \eta(\mathbf{x},t),
\end{equation}
with $\eta$ a Gaussian white noise of zero mean and unit variance, and $\zeta=\sqrt{2D}\,(i\nabla)^{a/2}\zeta'$.
Finite $\xi$ and $\tau$ imply spatiotemporal correlations in the noise $\zeta$, thereby breaking detailed balance in \cref{phieq} through a mismatch with the damping coefficient, which has no memory, and thus drives the system out of equilibrium \cite{fodor2016HowFar,maggi2022CriticalActive}.
Equivalently, \cref{phieq,OUeq} describe a non-reciprocal system with unidirectional coupling between $\phi$ and $\zeta'$, which violates the fluctuation-dissipation theorem (FDT) \cite{fruchart2026NonreciprocalManybody}.
This setup provides a minimal active generalization of the Brazovskii problem, corresponding to a finite-wavelength (type-$I_s$) instability in the classification of Cross and Hohenberg \cite{cross1993PatternFormation}.

\section{Self-consistent Hartree theory}
\label{sec:theory}
The effects of fluctuations in the Langevin-type equation (\ref{phieq}) can be analyzed using the response field formalism \cite{tauberCriticalDynamics,janssen1989NewUniversal} (see also \cref{app:msr}). Following this, we obtain the Gaussian (bare) response propagator,
\begin{equation}
    \label{bareG}
    G_{\mathbf{q}0}(t,t') = e^{-\Delta_{q0}|t-t'|}\theta(t-t'),
\end{equation}
where the bare relaxation rate, $\Delta_{q0} = Dq^a\bigl[r_0 + \kappa(q^2 - q_0^2)^2\bigr]$, with $\theta$ representing the Heaviside theta function. And the bare correlation function,
\begin{equation}
    \label{bareC}
    C_{\mathbf{q}0}(t,t')
    = \int\limits_{t_1=0}^{t}\int\limits_{t_2=0}^{t'}G_{\mathbf{q}0}(t,t_1)M_{\mathbf{q}}(t_1,t_2)G_{\mathbf{q}0}(t',t_2),
\end{equation}
with the noise correlation $\left<\zeta(\mathbf{q},t)\zeta(\mathbf{-q},t')\right> = M_\mathbf{q}(t,t') = \frac{Dq^a}{\tau\lambda_q}\exp\left[-\frac{\lambda_q|t-t'|}{\tau}\right]$, where $\lambda_q \equiv 1+\xi^2q^2$ denotes the spatial kernel of the colored noise, a shorthand we use throughout.

Fluctuation corrections to the bare propagator are incorporated through the Dyson equation \cite{bouchaud1996ModecouplingApproximations},
\begin{align}
    \label{dyson}
    G_\mathbf{q}(t,t') & =  G_{\mathbf{q}0}(t,t')\nn \\
    & +\int\limits_{t_1=t'}^{t}\int\limits_{t_2=t'}^{t_{1}}G_{\mathbf{q}0}(t,t_{1})\;\Sigma_\mathbf{q}(t_{1},t_{2})G_\mathbf{q}(t_{2},t'),
\end{align}
where one-loop self-energy is given by
\begin{equation}
    \label{selfe}
    \Sigma_\mathbf{q}(t_1,t_2) = -Dq^a\frac{u}{2}\int_{\mathbf{p}}C_{\mathbf{p}0}(t_1,t_2) + \mathcal{O}(u^2),
\end{equation}
with $\mathbf{p}$ denoting the internal momentum.
From \cref{dyson,selfe}, at early times, we find the time-dependent renormalized mass parameter $r(t)$:
\begin{equation}
    \label{rteq}
    r(t) = r_0 + \frac{u}{2}\int_{\mathbf{q}}C_{\mathbf{q}}(t,t),
\end{equation}
with the unsteady equal-time correlation function,
\begin{align}
    \label{Ctt}
    C_{\mathbf q}(t,t) &= \frac{D q^{a}}{\lambda_q \Delta_q
    \left(\tau^{2}\Delta_q^{2}-\lambda_q^{2}\right)}
    \Bigl[\lambda_q\!\left(e^{-2\Delta_q t}-1\right)
    \\ \nn
    &\qquad\quad +\tau\Delta_q
    \left(1+e^{-2\Delta_q t} 
    -2e^{-\left(\Delta_q+\lambda_q/\tau\right)t}\right)\Bigr].
\end{align}
Note that \cref{rteq} with \cref{Ctt} have been made self-consistent by replacing the bare relaxation rate $\Delta_{q0}$ in \cref{bareC} by its renormalized value $\Delta_q = D q^a\bigl[r + \kappa(q^2 - q_0^2)^2\bigr]$ \cite{frey1996ModecouplingRenormalization}. The apparent pole in \cref{Ctt} at $\Delta_q\tau=\lambda_q$ is removable, as the numerator vanishes at the same point; consequently, only the numerical evaluation requires care near this manifold.

At short times, \cref{Ctt} satisfies $\lim_{t \to 0} C_{\mathbf{q}}(t,t) = 0$, implying through \cref{rteq} that the mass parameter receives no Hartree-level (self-consistent one-loop) correction in this regime. Consequently, following quench below the mean-field transition point, the system initially exhibits unstable growth toward the ordered phase, unaffected by activity-induced corrections.
Indeed, in this early-time regime, the isotropic phase is genuinely linearly unstable for quenches below the mean-field transition ($r_0<0$): 
a spinodal does exist transiently, with the $|\mathbf q|\simeq q_0$ shell modes growing at rate $|\Delta_{q0}| \simeq Dq_0^a|r_0|$ \cite{gross2000KineticsOrdering}. On the other hand, the absence of a spinodal discussed below is a statement about the steady state, where the accumulated fluctuation corrections restore $r>0$ and convert the unstable growth into nucleation-limited kinetics.

At long times ($t\to\infty$), however, \cref{rteq} becomes
\begin{equation}
    \label{selfconstr}
    r = r_0 + \frac{u}{2}\int_{\mathbf{q}}C_{s}(\mathbf{q}),
\end{equation}
with the steady state correlation function,
\begin{equation}
    \label{Csteady}
    C_s(\mathbf{q})
    = \frac{Dq^a}{\lambda_q\Delta_q(\lambda_q+\tau\Delta_q)}.
\end{equation}

In the equilibrium limit, $\tau,\xi \to 0$, \cref{Csteady} recovers the
equilibrium equal-time correlator, $C^{\text{eq}}_s(\mathbf{q}) = D q^a/\Delta_q$, as expected; written out, $C^{\text{eq}}_s(\mathbf{q}) = \bigl[r+\kappa(q^2-q_0^2)^2\bigr]^{-1}$, with the mobility factor $Dq^a$ cancelling between noise and damping as detailed balance requires. 
Although the perturbative analysis above was carried out in the time domain and the steady-state limit was subsequently obtained by taking $t\to\infty$, an equivalent calculation performed directly in frequency space yields the same self-consistent equations, \cref{selfconstr}, together with the steady-state correlator \cref{Csteady}.
Furthermore, since the integral in \cref{selfconstr} is dominated by modes with $|\mathbf q| \simeq q_0$, the exponent $a$ does not qualitatively affect the results in the small-$r$ regime. In particular, for the conserved case ($a=2$), the extra $q^2$ factor can be absorbed into a redefinition of $D$, while inducing only a negligible fractional displacement of the dominant shell of order $r/q_0^4$
\cite{gross2000KineticsOrdering}.
The signature of the active effects can be read off directly from \cref{Csteady}. Writing $C_s(\mathbf q) = T_{\mathrm{eff}}(\mathbf q)\,C^{\text{eq}}_s(\mathbf q)$ identifies a mode-dependent ``effective temperature'':
\begin{equation}
    \label{Teff}
    T_{\mathrm{eff}}(\mathbf q) = \frac{1}{\lambda_q(\lambda_q+\tau\Delta_q)} \le 1,
\end{equation}
which decreases monotonically with both $\tau$ and $\xi$ (the equilibrium limit of bath temperature is unity in our units). 
Since every mode therefore remains at or below the equilibrium temperature, the fluctuation correction in \cref{selfconstr} can only be suppressed. 
Although many perturbations modify the static correlator, colored noise does so with a definite sign and, for finite $\tau$ or $\xi$, in a way that cannot be reproduced by an equilibrium model at a single temperature: the effective temperature $T_{\mathrm{eff}}$ becomes mode dependent, with FDT violations emerging at frequencies $\gtrsim \lambda_q/\tau$ \cite{maggi2022CriticalActive}.
On the shell, however, $\lambda_q\simeq\lambda_{q_0}$ is essentially constant, so the $\xi$-effect is a common temperature below the equilibrium value; the residual mode dependence at finite $\tau$ enters through $\Delta_q$, which varies across the shell. We now analyze separately the roles of the two activity parameters, $\tau$ and $\xi$, in the steady state.
\begin{figure}[t]
    \hspace*{-0.5cm}
    \includegraphics[width=\columnwidth]{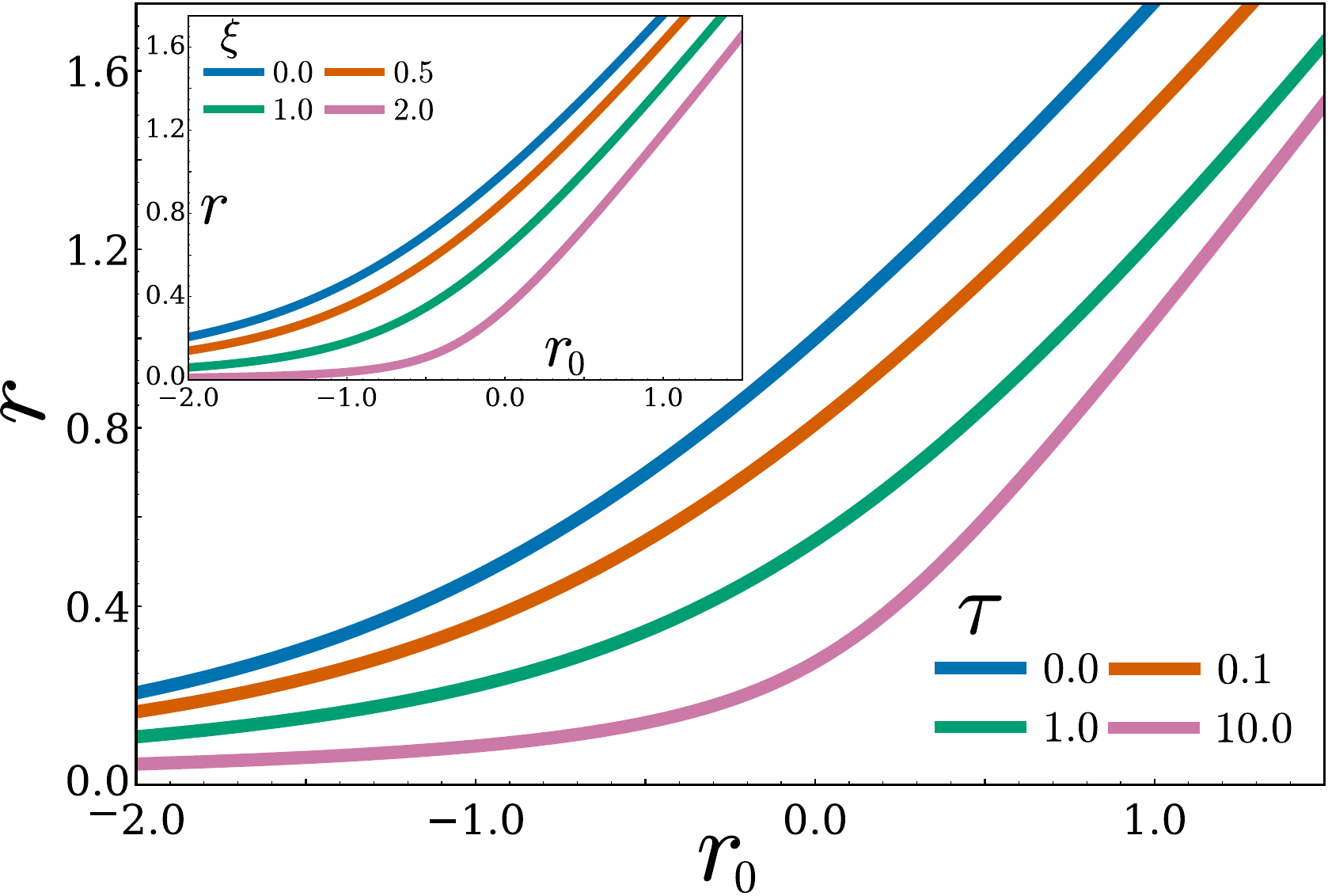}
    \caption{Renormalized mass parameter $r$ as a function of $r_0$ for different activity strengths $\tau$, including the equilibrium ($\tau=0$) curve for comparison. Increasing $\tau$ suppresses fluctuation corrections, reduces $r$, and shifts the transition to higher temperatures. The renormalized mass $r$ remains strictly positive, indicating the absence of a spinodal instability. Inset: Dependence on activity $\xi$.}
    \label{solutionplot}
\end{figure}

\subsection{Effect of the persistence time $\tau$}
To isolate the effect of the persistence time $\tau$ and simplify the notation, we set $\xi=0$ and $D=1$ in \cref{Csteady}.
We further absorb into a redefinition of $u$ the dimension-dependent, nonsingular prefactor of the integral in \cref{selfconstr}, evaluated around $|\mathbf{q}|\approx q_0$, namely $q_0^{d-2}/[2^{d+1}\pi^{d/2-1}\Gamma(d/2)\kappa^{1/2}]$
(for $a=0$).
Explicitly, since the integrand of \cref{selfconstr} is sharply peaked on the shell, we write $|\mathbf q| = q_0+\epsilon$ with $\epsilon\ll q_0$, so that $\Delta_q \simeq r+4\kappa q_0^2\epsilon^2$; the angular integration then gives a nonsingular factor $S_d\,q_0^{d-1}/(2\pi)^d$, with $S_d = 2\pi^{d/2}/\Gamma(d/2)$, while the $\epsilon$-integral extended over the real line. The partial-fraction identity $\frac{1}{\Delta_q(1+\tau\Delta_q)} = \frac{1}{\Delta_q} - \frac{\tau}{1+\tau\Delta_q}$
splits the radial integral into two pieces,
\begin{equation}
    \begin{aligned}
    \int_{-\infty}^{\infty}\frac{d\epsilon}{r+4\kappa q_0^2\epsilon^2} &= \frac{\pi}{2q_0\sqrt{\kappa}}\;r^{-1/2} \\
    \int_{-\infty}^{\infty}\frac{\tau\,d\epsilon}{1+\tau r+4\tau\kappa q_0^2\epsilon^2} &= \frac{\pi}{2q_0\sqrt{\kappa}}\;\tau^{1/2}(1+\tau r)^{-1/2}
    \end{aligned}
\end{equation}
which carry the same prefactor; collecting it with the angular factor and the $u/2$ of \cref{selfconstr} yields the above redefinition of $u$.
So, we find the self-consistent equation for the renormalized mass parameter $r$ as
\begin{equation}
    \label{selfconstr2}
    r = r_0 + u\left[r^{-1/2} - \tau^{1/2}(1+\tau r)^{-1/2}\right].
\end{equation}

\cref{selfconstr2} admits no finite solution for $r=0$, both in absence ($\tau=0$) and presence ($\tau>0$) of activity. Thus, at steady state, the disordered state remains metastable, and the transition proceeds via nucleation, as in the equilibrium Brazovskii scenario \cite{brazovskii1995PhaseTransition}.

However, as shown in \cref{solutionplot}, increasing $\tau$ suppresses the fluctuation-induced renormalization of the mass, resulting in smaller values of $r$ and shifts the transition to higher temperatures.
A similar effect of activity is evident at $\mathcal{O}(\tau)$ in the mean-field level, where the static properties are governed by an effective Boltzmann measure \cite{fodor2016HowFar} (see \cref{app:effF}).
Activity therefore weakens the Brazovskii mechanism, consistent with the suppression of fluctuation effects observed in shear-driven systems \cite{cates1989RoleShear} and related active systems \cite{maitra2020EnhancedOrientational}. In contrast to shear-driven systems \cite{cates1989RoleShear}, however, activity alone does not induce a spinodal instability of the isotropic phase.
The distinction from shear is worth emphasizing. Shear advects fluctuations anisotropically through the critical shell, cutting off the temporal integral that defines the steady-state correlator; this effect is nonanalytic in the shear rate and removes the $r^{-1/2}$ singularity, restoring the spinodal below the mean-field transition temperature for sufficiently strong shear and selecting a lamellar phase with wavevector normal to both the flow velocity and its gradient \cite{cates1989RoleShear,marques1990HexagonalLamellar}. The isotropic colored noise has no such cutting-off effect: it preserves the full degeneracy of the shell, and $T_{\mathrm{eff}}(q_0)\to\lambda_{q_0}^{-2}>0$ as $r\to0$ from \cref{Teff}, so the $r^{-1/2}$ singularity survives with a reduced amplitude at any finite activity, and no spinodal appears. So, at the level of the statics the $\xi$-effect is indeed a detailed-balance-respecting reduction of the noise strength ($u\to\tilde u$, below \cref{selfconstr_xi}); finite $\tau$ is not, since $T_{\mathrm{eff}}$ then varies across modes near the critical shell.

\subsection{Effect of the correlation length $\xi$}
Setting $\tau \to 0$ in \cref{Csteady} and employing the same rescaling as before to ease the notation, we obtain the following self-consistent equation:
\begin{equation}
    \label{selfconstr_xi}
    r = r_0 + \tilde{u} r^{-1/2},\quad \tilde{u} \equiv u/\lambda_{q_0}^2.
\end{equation}

Finite correlation length $\xi$ thus only modifies the effective interaction strength, $u \to \tilde{u}$, without altering the qualitative structure of the equilibrium solution, so the transition still remains first order. However, increasing $\xi$ suppresses fluctuation corrections, reduces $r$ (as shown in the inset of \cref{solutionplot}), and renders the system increasingly weakly first order.

Two limits are of particular interest.  In the limit of small activity ($\tau,\xi \to 0$), the colored noise reduces to white noise, and the equilibrium results are recovered (e.g, in \cref{selfconstr2,selfconstr_xi}). In contrast, for large activity ($\tau,\xi \to \infty$), fluctuation corrections vanish, and the system crosses over to mean-field behavior.
A similar suppression of fluctuation corrections to the four-point vertex by activity is observed at steady state (see \cref{app:vertex}); notably vertex corrections can drive the quartic coupling $u$ negative, thereby requiring stabilization via a $\phi^6$ term and ultimately yielding a fluctuation-induced first-order transition \cite{brazovskii1995PhaseTransition,fredrickson1989KineticsMetastable}. 

\section{Ordered phase}
\label{sec:ordered}
Here we derive the equation of state of the ordered phase.
We expand about the new potential minimum by decomposing the order parameter field as $\phi = \bar{\phi} + \psi$, where $\psi$ represents fluctuations away from the ordered state $\bar{\phi}$. Now, the renormalization of the mass term receives contributions from two diagrams: a one-loop tadpole diagram, analogous to that in the disordered phase (with $\psi$ replacing $\phi$), and a tree-level diagram containing two insertions of the ordered field $\bar{\phi}$ \cite{brazovskii1995PhaseTransition}.
Note that, in the active system, fluctuation corrections differ from their passive counterparts only through diagrams containing integrals over the correlation function \cref{bareC}.
Considering a lamellar state $\bar{\phi} = 2A \cos(\mathbf{q}_0\cdot \mathbf{x})$ with amplitude $2A$ and conjugate external field $h$, we obtain (following \cite{brazovskii1995PhaseTransition}), at steady state (for $\xi = 0$),
\begin{subequations}
    \label{ordered_eq}
\begin{align}
& h = rA-\frac{u}{2}A^3, \\
r = r_0 + uA^2 + & u\left[r^{-1/2}-\tau^{1/2}(1+\tau r)^{-1/2}\right].
\end{align}
\end{subequations}
Here and below, $h$ and $A$ denote reduced variables rescaled together with $u$, such that the same rescaled coupling $u$ in \cref{selfconstr2} multiplies both the tree-level term $uA^2$ and the Hartree correction \cite{brazovskii1995PhaseTransition,hohenberg1995MetastabilityFluctuationdriven}.

For $h=0, A\neq0$, we have
\begin{equation}
    \label{selfconstr3}
    -r_0 = r + u\left[r^{-1/2}-\tau^{1/2}(1+\tau r)^{-1/2}\right].
\end{equation}

For small $\tau$, \cref{selfconstr3} admits real solutions for $r$ provided $-r_0 \ge r_c$, where $r_c = 3(u/2)^{2/3} - u\,\tau^{1/2} + \mathcal{O}(\tau^{3/2})$.
Here, $r_c$ defines limits of metastability of the ordered phase (or, superheating limit \cite{chaikin2012PrinciplesCondensed}):
the condition $-r_0 = r_c$ marks the onset of the ordered state, with bistable amplitudes $A_c = \pm(2r_c/3u)^{1/2}$ \cite{komura2008DynamicalBrazovskii}. Note that, increasing $\tau$ lowers $r_c$ relative to its equilibrium value $r_c^{\mathrm{eq}} = 3(u/2)^{2/3}$.

In the small $\tau$ limit, the difference between the effective potentials of the ordered/lamellar ($l$) and disordered ($d$) states reads
\begin{align}
    \label{DelPhi}
    \Delta \Phi  = \int\limits_0^A h\frac{\partial\bar{\phi}}{\partial A'} dA'
    & = -\frac{r_l^2 + r_d^2}{2u}+ (r_l^{1/2}-r_d^{1/2})\nn \\
    &-\frac{\tau^{3/2}}{4}(r_l^2 - r_d^2)
    + \mathcal{O}(\tau^{5/2}).
\end{align}

A scaling analysis of \cref{DelPhi}, together with \cref{selfconstr2,selfconstr3}, reveals the existence of a characteristic threshold $r_c' \sim u^{2/3} > r_c$, with a subleading negative corrections of $\mathcal{O}(u\tau^{1/2})$ such that $\Delta \Phi < 0$ for $-r_0 > r_c'$.
So, the ordered state remains metastable for $r_c' > -r_0 > r_c$ and becomes thermodynamically favorable for $-r_0 > r_c'$; the disordered state, in contrast, is metastable throughout.
Finally, for $\xi \neq 0$ and $\tau \to 0$, the analysis reduces to the equilibrium form [e.g., \cref{ordered_eq} without the $\tau$ term, or equivalently the first line of \cref{DelPhi}], but with the modified interaction strength $\tilde{u}$. In this limit, one again finds $\Delta \Phi < 0$, confirming the robustness of the fluctuation-induced first-order transition, with finite $\xi$ effectively suppressing fluctuation effects, shifting the critical parameters, and thereby stabilizing the ordered phase.
\begin{figure}[t]
    \hspace*{-0.5cm}
    \includegraphics[width=\columnwidth]{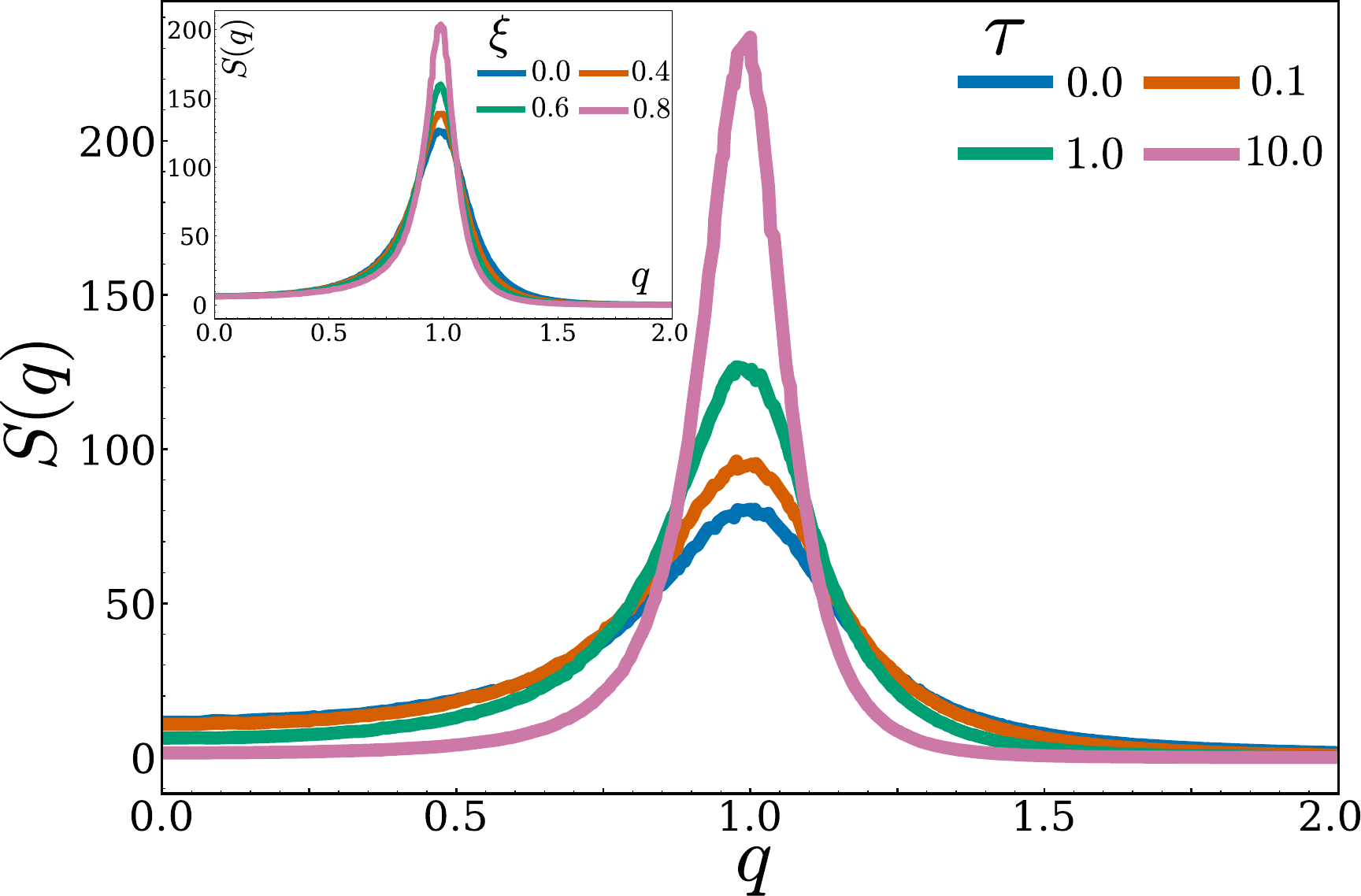}
    \caption{Static structure factor $S(q)$ as a function of wave number $q$ for increasing activity $\tau$, showing an enhanced and sharpened dominant peak. Inset: $S(q)$ for increasing $\xi$. Parameters: $r_0=-0.5, u=6.0, \kappa=q_0=1.0, \tilde{D}=0.5$ (see \cref{app:dns} for its definition), $\xi=a=0$; $\tau=1.0$ in the inset.}
    \label{sk2d}
\end{figure}

\section{Numerical simulations}
\label{sec:dns}
To corroborate the theoretical predictions, we perform direct numerical simulations (DNS) of \cref{phieq,OUeq} using a pseudospectral scheme with periodic boundary conditions in two spatial dimensions (see \cref{app:dns} for details). \cref{sk2d} shows the time-averaged steady-state static structure factor (SF) for different activity strength, demonstrating enhanced ordering with increasing activity. The SF peak height as function of activity is quantified in \cref{app:sqpeak}. Note that these simulations, at fixed system size, corroborate the trends of the self-consistent theory in \cref{sec:theory}; establishing the order of the transition and its weakening with activity would require finite-size scaling of the susceptibility and Binder cumulant, together with hysteresis and nucleation statistics, a substantially larger computational campaign that we defer to our future work.

Moreover, fluctuations are known to strongly affect both passive \cite{toner1981SmecticCholesteric} and active lamellar (smectic) phases \cite{adhyapak2013LiveSoap,julicher2022BrokenLiving} in two dimensions, particularly through long-wavelength modes and dislocations. Consequently, the ordered state observed in our noisy two-dimensional DNS is more appropriately interpreted as a nematic phase with orientational order of the local layer normal \cite{toner1981SmecticCholesteric}, rather than a true smectic. Within this framework, our numerical results do not unambiguously demonstrate activity-induced enhancement of smectic order in two dimensions; instead, they are consistent with activity suppressing long-wavelength fluctuations and thereby stabilizing the nematic order (see \cref{app:nematic}).

\section{Summary and outlook}
\label{sec:summary}
To summarize, we investigated how activity alters fluctuation-induced first-order transitions in the Brazovskii universality class, examining both the approach to stationarity and the properties of the resulting steady state.
We showed that activity does not modify the early-time mean-field dynamics, since fluctuation corrections remain negligible in this regime \cite{gross2000KineticsOrdering}.
At longer times, however, fluctuations become essential and render the transition first order, leading to a persistent metastable disordered phase and ordering dynamics controlled by nucleation.
Increasing activity suppresses the fluctuations underlying the Brazovskii mechanism, shifts the transition to higher temperatures, and progressively weakens its first-order character without inducing a spinodal instability of the isotropic phase.
In the infinite-activity limit, fluctuation corrections vanish, and the system crosses over to mean-field behavior.
In the end, then, the statics of the transition reported here are primarily an activity-induced renormalization of the passive theory; the genuinely nonequilibrium features, the mode dependence of $T_{\mathrm{eff}}$ and the FDT violations, reside in the dynamics and response.
Notably, following a deep quench below the mean-field transition, the dynamics initially exhibit spinodal-like growth before ultimately crossing over to nucleation-controlled steady-state kinetics, whose detailed analysis we defer to our future work (for the passive case, see, e.g., \cite{gross2000KineticsOrdering}).

Possible physical realizations relevant to our theory include systems exhibiting finite-wavelength ordering, such as microphase-separated morphologies, under nonequilibrium conditions. Equilibrium colloidal systems with competing interactions provide well-established realizations of stripe and cluster microphases \cite{imperio2006MicrophaseSeparation}, whose fluctuation-driven ordering may be modified by activity. Related phenomenology also arises in living liquid crystals \cite{zhou2014LivingLiquid} and in active block-copolymer assemblies. The latter may be realized either through passive copolymers immersed in active baths \cite{kaiser2014UnusualSwelling} or through assemblies composed of intrinsically active monomers \cite{jiao2023ConfigurationDynamics,winkler2020PhysicsActive}.
Note that, in block copolymer melts (\cref{diblock}), incompressibility and chain-connectivity-induced long-range interactions stabilize finite-wavelength patterns \cite{leibler1980TheoryMicrophase,ohta1986equilibrium}, placing their active counterparts in the weak-segregation limit \cite{gross2000KineticsOrdering,qi1996KineticPathways} within our non-conserved ($a=0$) class.

\section{Acknowledgments}
\label{sec:acknowledge}
The author gratefully acknowledges Sriram Ramaswamy for many insightful discussions and for a careful reading of the manuscript. The author also thanks Ananyo Maitra and K. Vijay Kumar for valuable comments and suggestions.


\appendix
\section*{Appendix}
\subsection{Dynamical action formalism}
\label{app:msr}
The dynamical action \cite{tauberCriticalDynamics}, expressed as a functional of the scalar field $\phi$ and the Martin--Siggia--Rose response field $\tilde{\phi}$, can be written as $\mathcal{A}[\tilde{\phi},\phi] = \mathcal{A}_0 + \mathcal{A}_{\mathrm{int}}$, where the bilinear part
\begin{align}
    \label{Aharm}
    \mathcal{A}_0[\tilde{\phi},\phi]
    &= \int_{\mathbf{x},t}
    \tilde{\phi}
    \Big[\partial_t + D(i\nabla)^a \big(r_0 + \kappa(\nabla^2+q_0^2)^2\big)\Big]
    \phi \nn\\
    &- \int_{\mathbf{x},\mathbf{x'},t,t'}
    D(i\nabla)^a \tilde{\phi}\,
    K_{\xi,\tau}(|\mathbf{x}-\mathbf{x'}|,|t-t'|)\,
    \tilde{\phi},
\end{align}
and the anharmonic part
\begin{equation}
    \label{Aanharm}
    \mathcal{A}_{int}[\tilde{\phi},\phi] = \frac{Du}{6}\int_{\mathbf{x},t}\tilde{\phi}(\mathbf{x},t)(i\nabla)^a\phi(\mathbf{x},t)^3.
\end{equation}

Mean values of operators over the noise history are expressed as functional integrals with weight $e^{-\mathcal{A}[\tilde{\phi},\phi]}$.
Although average should, in principle, be taken over initial conditions at early times, their influence becomes negligible near the critical point at large length scales, up to a sub-leading correction \cite{janssen1989NewUniversal}. All response and correlation functions can be obtained as functional derivatives
of the generating functional:
\begin{equation}
    \label{genfunc}
    \mathcal{Z} = \ln\int\mathcal{D}[i\tilde{\phi},\phi]e^{-\mathcal{A}[\tilde{\phi},\phi]}\exp{\int_{\mathbf{x},t}\left(h\phi  + \tilde{h}\tilde{\phi}\right)},
\end{equation}
with respect to the conjugate fields $h(\mathbf{x},t)$ and $\tilde{h}(\mathbf{x},t)$.
Using $\mathcal{A}_0$ in \cref{genfunc}, we obtain the bare response function $G_0(t,t') = \langle \tilde{\phi}(t)\,\phi(t') \rangle_0$ and the bare correlation function $C_0(t,t') = \langle \phi(t)\,\phi(t') \rangle_0$, consistent with the expressions derived in the main text \cite{janssen1989NewUniversal,bouchaud1996ModecouplingApproximations}.
The fourth-order interaction term in \cref{Aanharm} can then be decoupled in a self-consistent manner as
\begin{equation}
    \frac{Du}{6}\langle\tilde{\phi}\phi\phi\phi\rangle_0 \rightarrow \frac{Du}{2}\langle\tilde{\phi}\phi\rangle_0\langle\phi\phi\rangle_0,
\end{equation}
which leads to a renormalization of the bare response propagator in \cref{dyson} and, consequently, a time dependent shift in the mass parameter in \cref{rteq}.

\subsection{Effective free-energy functional}
\label{app:effF}
Adiabatically eliminating the noise $\zeta$ for $\xi=0$ \cite{fodor2016HowFar,maitra2020EnhancedOrientational,ryter1981BrownianMotion}, we find the effective free-energy functional at $\mathcal{O}(\tau)$:
\begin{equation}
    \begin{aligned}
    \tilde F[\phi] &= F[\phi] + \tau D\int_{\mathbf x}\left[\frac12\left(\frac{\delta F}{\delta\phi}\right)^{2}
    - \frac{\delta^2F}{\delta\phi^2}\right] \\
    &= \int_{\mathbf x}\Bigg[\frac12\bigl[r_0+\tau D(r_0^2-u)\bigr]\phi^2 \\
    &\qquad + \frac{u}{4!}\bigl(1+4\tau D r_0\bigr)\phi^4 + \frac{\tau D u^2}{72}\phi^6 \\
    &\qquad + \frac{\kappa}{2}\bigl(1+2\tau D r_0\bigr) \bigl[(\nabla^2+q_0^2)\phi\bigr]^2 \\ 
    &\qquad + \frac{\tau D\kappa^2}{2}\bigl[(\nabla^2+q_0^2)^2\phi\bigr]^2 \\ 
    &\qquad + \frac{\tau D\kappa u}{6}\,\phi^3(\nabla^2+q_0^2)^2\phi\Bigg].
\end{aligned}
\end{equation}
As evident from the first term in the integral, for small
$r_0$ the effective coefficient of $\phi^2$ is reduced, corresponding to a shift of the mean-field critical point from $0$ to $\tau u D$, i.e., already at $\mathcal O(\tau)$ activity favors ordering, consistent with the suppression of fluctuation effects found above.
In the complementary limit ($\tau\to0$, $\xi\neq0$), the noise is reduced by $\lambda_q^2$, as discussed in \cref{sec:theory}.
Written as a correction to $F$, this gives a gradient term $\sim\xi^2\int_{\mathbf x}(\nabla\frac{\delta F}{\delta\phi})^2$ for $\xi\ll1$ \cite{garcia-ojalvo1994ColoredNoise}, which cannot directly rescale the mean-field quadratic coupling. Thus, the critical point remains at $r_0=0$, while the suppression enters only through the one-loop correction, with a detailed-balance-respecting reduction of the noise at the shell $|\mathbf{q}|\approx q_0$, $u\to\tilde u$ in \cref{selfconstr_xi}, shifting the transition to higher temperatures.

\subsection{Vertex correction}
\label{app:vertex}
At steady state, as in the equilibrium case \cite{brazovskii1995PhaseTransition}, the four-point vertex $\Gamma^{(4)}(\mathbf{q},-\mathbf{q},\mathbf{q}',-\mathbf{q}') \equiv \Gamma^{(4)}(\mathbf{q},\mathbf{q}')$ receives three $\mathcal{O}(u^2)$ contributions: one from the ladder diagram for $\mathbf{q} \neq \pm \mathbf{q}'$, and two additional contributions arising in the special cases $\mathbf{q} = \pm \mathbf{q}'$, where $\mathbf{q}$ and $\mathbf{q}'$ denote the external momenta. Using resummation, we get
\begin{equation}
    \label{g41}
    \Gamma^{(4)}(\mathbf{q},\mathbf{q}')
    = u - u\Pi\,\Gamma^{(4)}
    = \frac{u}{1+u\Pi},
\end{equation}
\begin{equation}
    \label{g42}
    \Gamma^{(4)}(\mathbf{q},\pm\mathbf{q}') = u + 2\left(\frac{u}{1+u\Pi}-u\right)  = \frac{u(1-u\Pi)}{1+u\Pi},
\end{equation}
where $\Pi$ denotes the loop integral over internal momenta.
The integral is dominated by modes with $|\mathbf{q}| \simeq q_0$, yielding, for $\xi=0$, the following expression in frequency space:
\begin{align}
    \label{Pi_int}
    \Pi & = \int_{\mathbf{q},\omega}C_{\mathbf{q}0}(\omega)G_{\mathbf{q}0}(\omega)\nn \\
    & \sim\int_{\mathbf{q},\omega}\frac{2Dq^a}{(\omega^2 + \Delta_{q0}^2)(1 +\tau^2\omega^2)(-i\omega+\Delta_{q0})} \nn \\
    & \sim\int_{\mathbf{q}}\frac{Dq^a(1+2\tau\Delta_{q0})}{\Delta_{q0}^2(1+\tau\Delta_{q0})^2}.
\end{align}

In the limit $\tau \to 0$, $\Pi$ in \cref{Pi_int} reduces to its equilibrium form \cite{brazovskii1995PhaseTransition}, as expected. 
In this case, $\Pi \sim r_0^{-3/2}$, which diverges as $r_0 \to 0$, ensuring a sign change of the effective quartic coupling near criticality through \cref{g42}, necessitating the inclusion of a stabilizing $\phi^6$ term and thereby enabling a fluctuation-induced first-order transition.
For finite $\tau$, however, the vertex corrections in \cref{g41,g42} are suppressed relative to their equilibrium counterparts and decrease monotonically with increasing $\tau$ values.
A similar suppression of fluctuation corrections arises for finite $\xi$ with $\tau\to0$, where the equilibrium form is recovered with the modified interaction strength $\tilde{u}$, as discussed in the main text.
In the limit $\tau \to \infty$, $\Pi \to 0$, indicating the complete suppression of vertex corrections, analogous to the behavior of the mass correction; for $\xi \to \infty$, the effective interaction strength $\tilde{u}$ vanishes identically.

The quartic coupling turning negative may appear paradoxical, so a clarification (following closely \cite{hohenberg1995MetastabilityFluctuationdriven}) is in order. Differentiating \cref{ordered_eq} with respect to $A^2$ for $\tau=\xi=0$ gives
\begin{equation}
    \label{drdA2}
    \frac{\partial r}{\partial A^2} = \frac{u}{1+\frac{u}{2}r^{-3/2}},
\end{equation}
which is precisely the resummed vertex (\ref{g41}) with $u\Pi = \frac{u}{2}r^{-3/2}$ inheriting nonanalytic character from the $r^{-1/2}$ integrand. 
Now eliminating $h$ from \cref{DelPhi}, $\Delta\Phi(A) = 2\int_0^Ah\,dA'$, using \cref{ordered_eq} and substituting $x = A'^2$ gives
\begin{equation}
    \label{PhiInt}
    \Delta\Phi(A) = \int_0^{A^2}\Bigl[r(x)-\frac{u}{2}x\Bigr]dx .
\end{equation}
Taylor expanding $r(x) = r_d+r_1x+r_2x^2+\mathcal{O}(x^3)$, with $r_d>0$ the disordered solution of \cref{selfconstr2}, and setting $g \equiv 1+\frac{u}{2}r_d^{-3/2}$, \cref{drdA2} and its derivative give $r_1 = u/g$ and $r_2 = \frac{3u^3}{8g^3}r_d^{-5/2}$, so that
\begin{equation}
    \label{PhiExp}
    \Delta\Phi(A) = r_dA^2 + \frac{u(2-g)}{4g}A^4 + \frac{u^3r_d^{-5/2}}{8g^3}A^6 + \mathcal{O}(A^8).
\end{equation}
From \cref{PhiExp}, the curvature at the origin, $\Delta\Phi''(0) = 2r_d$, is strictly positive for every $r_0$, so the disordered state never loses local stability. The quartic coefficient turns negative for $g>2$, which is exactly where $u\Pi=1$ in \cref{g42} changes sign. So the conventional shorthand phrasing, that the vertex turns negative \cite{brazovskii1995PhaseTransition} and a $\phi^6$ term is needed to stabilize the free-energy functional \cite{fredrickson1989KineticsMetastable}, encodes the first-order character of the transition.

\subsection{DNS details}
\label{app:dns}
We employ an exact time updating formula for the colored noise $\zeta'$ describing Ornstein--Uhlenbeck process (for $\tau\neq0$) as \cite{gillespie1996ExactNumerical}
\begin{equation}
    \zeta'(\mathbf{q}, t+\Delta t)
    =
    e^{-\lambda_q\frac{\Delta t}{\tau}}
    \zeta'(\mathbf{q}, t)
    +
    \sqrt{
    \frac{1 - e^{-2\lambda_q\frac{\Delta t}{\tau}}}{2\tau \lambda_q(\Delta x)^d}
    }
    \eta(\mathbf{q}, t),
\end{equation}
where $\lambda_q$ is defined in \cref{sec:model} and $\eta$ is the Fourier transform of Gaussian white noise with zero mean and unit variance.
The time evolution of the order parameter $\phi$ is computed using a pseudospectral method, with the dynamical equation in Fourier space given by:
\begin{equation}
    \label{genpde}
    \partial_{t}\phi(\mathbf{q},t) = \mathcal{L}(\mathbf{q})\phi(\mathbf{q},t)+\mathcal{N}[\{\phi(\mathbf{q},t)\}]+\zeta(\mathbf{q},t),
\end{equation}
where the linear operator $\mathcal{L}(\mathbf{q})=-q^a(r_0+\kappa(q^2-q_0^2)^2)$, nonlinear term $\mathcal{N}[\{\phi(\mathbf{q},t)\}] = -uq^a[\phi^3]_{\mathbf{q}}$ and the noise term is $\zeta(\mathbf{q},t)=(2\tilde{D}q^{a})^{1/2}\zeta'(\mathbf{q},t)$. Note that \cref{genpde} carries unit mobility, so $\tilde D$ plays the role of a temperature, or noise strength.
Because the linear operator $\mathcal{L}(\mathbf{q})$ is stiff (due to $q^{4+a}$ term), we employ a second-order implicit-explicit (IMEX) scheme \cite{ascher1995ImplicitExplicitMethods} for time stepping, together with $2/3$ de-aliasing rule to handle the $\phi^3$ nonlinearity. Simulations for the white-noise limit ($\tau,\xi \to 0$) were carried out by directly incorporating stochastic fluctuations into \cref{genpde}.

The computational domain comprises a $512^2$ grid points $(R^2)$ over a square domain of linear extent $32\pi~(L)$, and the system is evolved in time using a time step of $\Delta t = 0.01$. Initial conditions are drawn from Gaussian distributions, $\phi_0 = 0.5\,\mathcal{N}(0,1)$ and $\zeta'_0 = 0.1\,\mathcal{N}(0,1)$, where $\mathcal{N}(0,1)$ denotes a normal distribution with zero mean and unit variance. The structure factor shown in \cref{sk2d} is computed after the system has reached a steady state and subsequently time averaged over 1000 uncorrelated configurations sampled from a single long simulation run for each set of parameters.

\subsection{Structure-factor peak}
\label{app:sqpeak}
\begin{figure}[t]
    \includegraphics[width=\columnwidth]{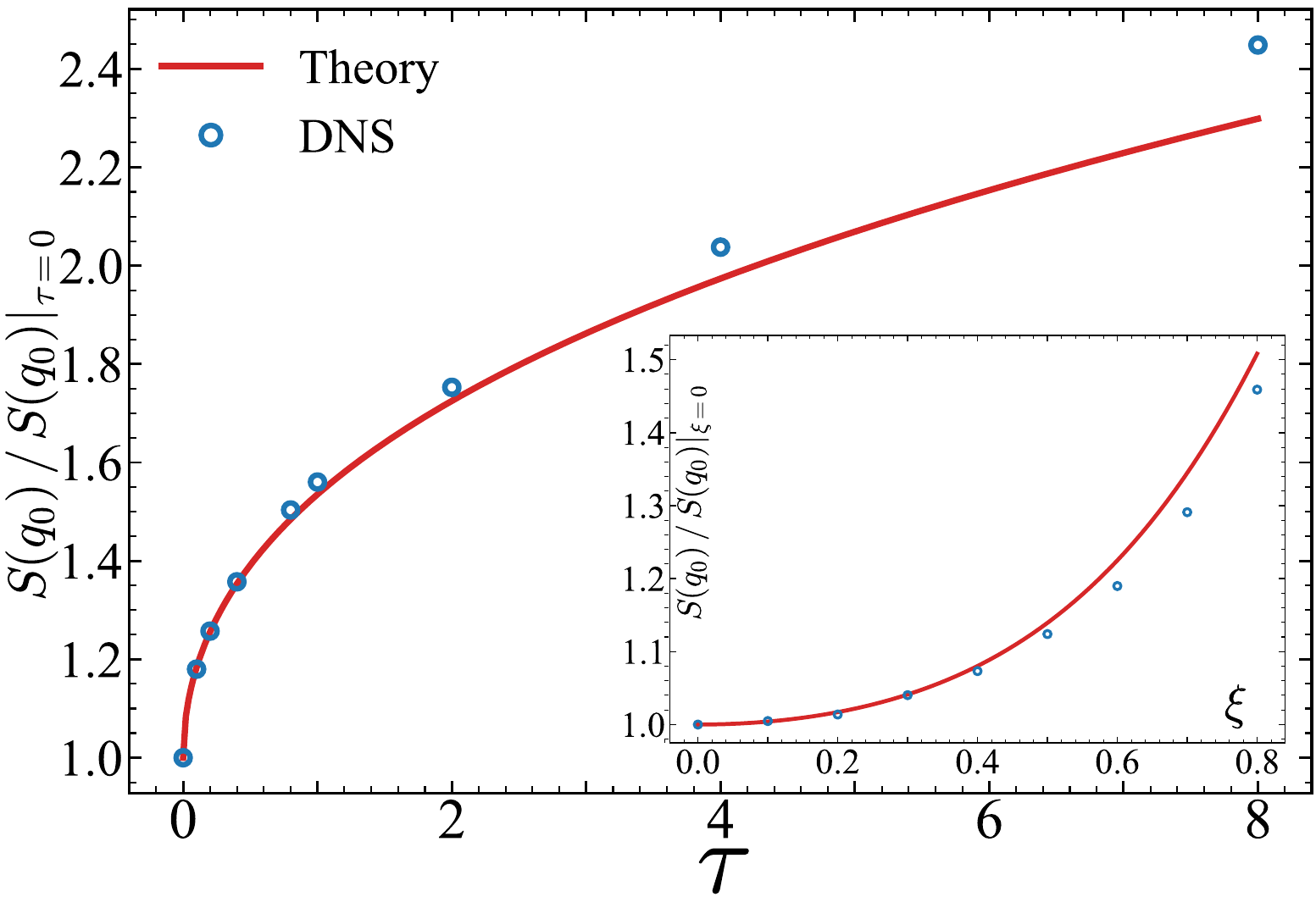}
    \caption{Structure-factor peak height $S(q_0)$, normalized by its zero-activity value, as a function of $\tau$ (inset: dependence on $\xi$). Parameters: $r_0=-0.5$, $u=6.0$, $\kappa=q_0=1.0$, $\tilde{D}=0.5$, and $a=0$; $\tau=1.0$ in the inset.}
    \label{peakfig}
\end{figure}

The steady-state structure factor is $S(\mathbf q) = C_s(\mathbf q)$, \cref{Csteady}, so the peak in \cref{sk2d} is fixed by the renormalized mass $r$. At the shell, $\Delta_{q_0} = Dq_0^ar$, and \cref{Csteady} gives
\begin{equation}
    \label{speak}
    S(q_0) = \left[\lambda_{q_0}\bigl(\lambda_{q_0}+\tau\Delta_{q_0}\bigr)\,r\right]^{-1} = T_{\mathrm{eff}}(q_0)\, r^{-1},
\end{equation}
with $T_{\mathrm{eff}}$ from \cref{Teff}, so the peak height measures the inverse renormalized mass weighted by the effective temperature, and reduces to $r^{-1}$ in the equilibrium limit. Note that \cref{speak} carries two competing effects of activity: $r^{-1}$ rises, as in \cref{solutionplot}, while $T_{\mathrm{eff}}(q_0)$ falls, as explained in \cref{sec:theory}. 
For $r_0<0$, the regime of \cref{sk2d}, the rise of $r^{-1}$ dominates, and the peak grows with both $\tau$ and $\xi$. At $\xi=0$, $S(q_0) = \left[r(1+\tau Dq_0^ar)\right]^{-1}$ with $r$ from \cref{selfconstr2}, increasing monotonically with $\tau$; 
at $\tau=0$, \cref{selfconstr_xi} gives $r\to u^2/(\lambda_{q_0}^4r_0^2)$ for $r\ll|r_0|$, so $S(q_0)\to(\lambda_{q_0}r_0/u)^2$ grows as $\lambda_{q_0}^2$. 
The trend reverses only for $r_0>0$, where $r\to r_0$ and the peak decays as $\lambda_{q_0}^{-2}$.
\cref{peakfig} compares \cref{speak} with the DNS result. For the inset, where $\tau$ and $\xi$ are both finite, $r$ is obtained by solving the full self-consistent equation \cref{selfconstr} with \cref{Csteady}.
The theory reproduces the enhanced peak observed in the DNS without any fitting, with deviations emerging only at larger activity, where the vertex correction \cref{Pi_int} neglected by the Hartree treatment grows in the self-consistent solution.

\subsection{Isotropic-to-nematic transition}
\label{app:nematic}
We define an orientational order parameter from the structure factor, analogous to the nematic tensor order parameter in liquid crystals \cite{deGennesProst1993}, adapted to Fourier-space and restricted to the dominant modes on a shell of radius $q_0$:
\begin{equation}
    Q_{ij} \equiv \frac{\sum_{q \in\mathcal{A}_{q_0}} S(q)\left(\hat{q}_i \hat{q}_j - \frac{1}{2}\delta_{ij}\right)} {\sum_{q \in\mathcal{A}_{q_0}} S(q)},
\end{equation}
where $\mathcal{A}_{q_0} = \{\mathbf{q}:|\mathbf{q}|\in[q_0-\Delta q, q_0+\Delta q]\}$ denotes a narrow annulus around $q_0$.

We quantify fluctuations of the orientational order using a fourth-order Binder-type cumulant constructed from the rotationally invariant nematic order parameter $Q=(Q_{xx}^2+Q_{xy}^2)^{1/2}$,
\begin{equation}
    \label{binder}
    U_4^Q \equiv 2-\frac{\langle Q^4\rangle}{\langle Q^2\rangle^2},
\end{equation}
where $\langle\cdot\rangle$ denotes steady-state time averaging.
The normalization in \cref{binder} is chosen such that $U_4^Q\to1$ in the ordered phase and $U_4^Q\to0$ in the disordered phase. In the latter regime, $Q_{xx}$ and $Q_{xy}$ behave approximately as independent zero-mean Gaussian variables, implying that the magnitude $Q$ follows a Rayleigh distribution (equivalently, a $\chi_2$ distribution).
\begin{figure}[t]
    \hspace*{-0.5cm}
    \includegraphics[width=\columnwidth]{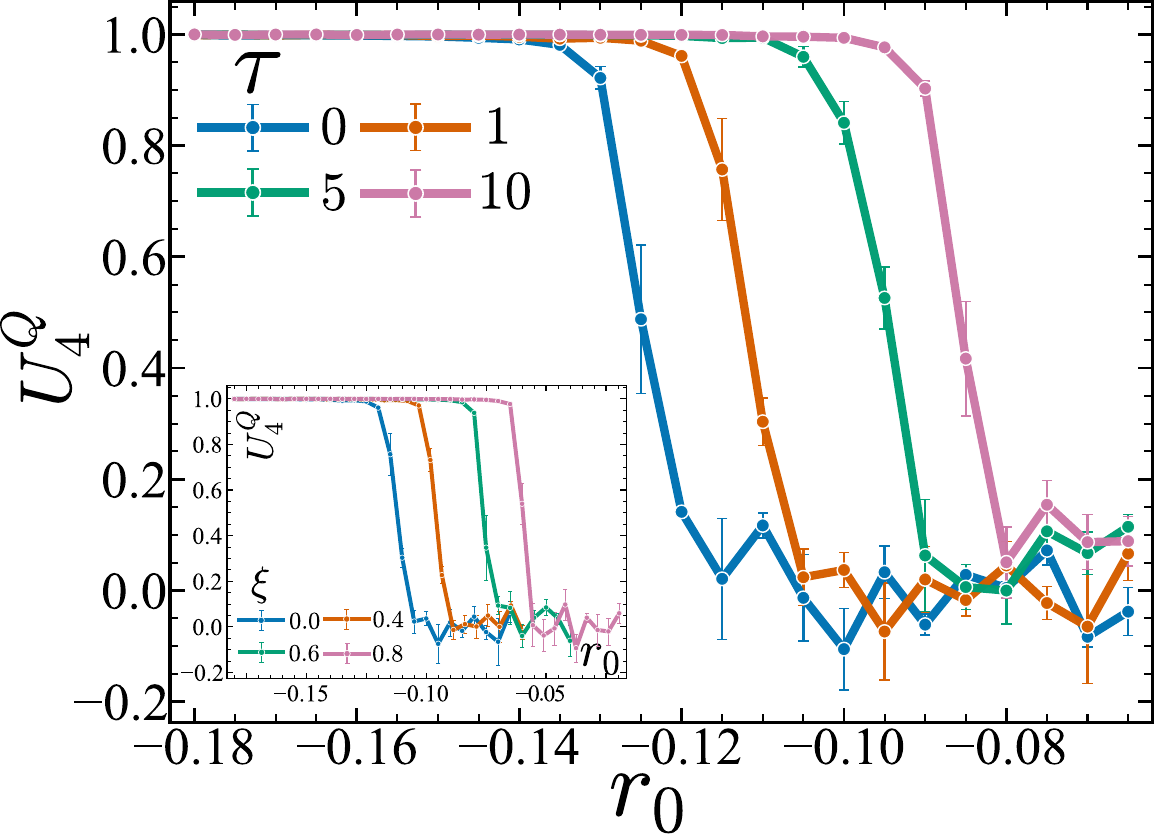}
    \caption{Binder cumulant $U_4^Q$ for the isotropic--nematic transition at different activity strengths $\tau$ (inset: dependence on $\xi$). Increasing activity shifts the transition to higher temperatures, indicating enhanced stability of the nematic phase.  Parameters: $u=6.0,  \kappa=q_0=1.0, \tilde{D}=0.02, \xi=a=0, L=16\pi, R=256$; $\tau=1.0$ in the inset.}
    \label{binder_c}
\end{figure}

Note from \cref{binder_c} that, at equilibrium ($\tau=0$), fluctuation corrections shift the transition temperature from the mean-field critical point at $0$ to a negative value. As activity increases, these fluctuation corrections are progressively suppressed, shifting the critical temperature toward higher values and thereby enhancing ordering.

\bibliography{colored}

\end{document}